\documentclass[aip,apl,amsmath,amssymb,reprint,]{revtex4-1}
\usepackage{graphicx}
\usepackage{epstopdf}
\usepackage{bm}
\usepackage{longtable}
\usepackage{float}
\usepackage{dcolumn}
\usepackage{bm}
\begin{document}
\preprint{AIP/123-QED}
\title[]{Reduction of light shifts in Ramsey spectroscopy with a combined error signal}
\author{M. Shuker}
\email{moshe.shuker@nist.gov}
\affiliation{National Institute of Standards and Technology, Boulder, Colorado 80305, USA}
\affiliation{University of Colorado, Boulder, Colorado 80309-0440, USA}
\author{J. W. Pollock}
\affiliation{National Institute of Standards and Technology, Boulder, Colorado 80305, USA}
\affiliation{University of Colorado, Boulder, Colorado 80309-0440, USA}
\author{R. Boudot}
\affiliation{FEMTO-ST, CNRS, 26 rue de l'\'epitaphe 25030 Besancon, France}
\affiliation{National Institute of Standards and Technology, Boulder, Colorado 80305, USA}
\author{V. I. Yudin}
\affiliation{Novosibirsk State University, ul. Pirogova 2, Novosibirsk, 630090, Russia }
\affiliation{Institute of Laser Physics SB RAS, pr. Akademika Lavrent'eva 13/3, Novosibirsk, 630090, Russia }
\affiliation{Novosibirsk State Technical University, pr. Karla Marksa 20, Novosibirsk, 630073, Russia }
\author{A. V. Taichenachev}
\affiliation{Novosibirsk State University, ul. Pirogova 2, Novosibirsk, 630090, Russia }
\affiliation{Institute of Laser Physics SB RAS, pr. Akademika Lavrent'eva 13/3, Novosibirsk, 630090, Russia }
\author{J. Kitching}
\affiliation{National Institute of Standards and Technology, Boulder, Colorado 80305, USA}
\affiliation{University of Colorado, Boulder, Colorado 80309-0440, USA}
\author{E. A. Donley}
\affiliation{National Institute of Standards and Technology, Boulder, Colorado 80305, USA}
\affiliation{University of Colorado, Boulder, Colorado 80309-0440, USA}
\begin{abstract}
Light-induced frequency shifts can be a key limiting contribution to the mid and long-term frequency instability in atomic clocks.  In this letter, we demonstrate the experimental implementation of the combined error signal interrogation protocol to a cold-atom clock based on coherent population trapping (CPT) and Ramsey spectroscopy. The method uses a single error signal that results from the normalized combination of two error signals extracted from two Ramsey sequences of different dark periods. The single combined error signal is used to stabilize the atomic clock frequency. Compared to the standard Ramsey-CPT interrogation, this method reduces the clock frequency sensitivity to light-shift variations by more than one order of magnitude. This method can be applied in various kinds of Ramsey-based atomic clocks, sensors and instruments.
\end{abstract}
\maketitle
\begin{figure}[t]
    \includegraphics[width=8.2cm]{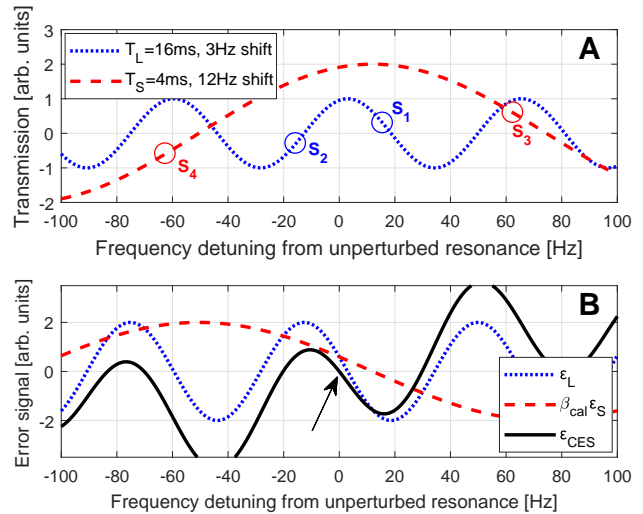}
    \caption{
    \label{Figure1_CES_principle}
    The principle of CES spectroscopy. A: Ramsey-CPT fringes with two different dark periods $T_L=16$ ms (dotted line) and $T_S=4$ ms (dashed line), assumed to be light-shifted by $3$ Hz and $12$ Hz respectively (neglecting all other shifts). Frequency jumps of $\pm \frac{1}{4T}$ from the clock target frequency (zero detuning) are shown by circles on the fringes ($S_1$, $S_2$ for the Ramsey fringe with $T_L$, $S_3$, $S_4$ for the Ramsey fringe with $T_S$). The Ramsey fringe with the longer dark-period has lower amplitude due to the excess decay. B: The error-signals $\varepsilon_L=S_1-S_2$, $\varepsilon_S=S_3-S_4$ (after normalization using $\beta_{cal}$) and $\varepsilon_{CES}=\varepsilon_L-\beta_{cal}\varepsilon_S$ extracted from the fringes in Fig. \ref{Figure1_CES_principle}.A  vs. the frequency detuning from the unperturbed resonance. It is evident that while $\varepsilon_L$ and $\varepsilon_S$ are non-zero, $\varepsilon_{CES}$ nullifies at the target clock frequency (zero detuning), as indicated by an arrow. This result is true for any value of shift that inversely depends on $T$. 
    }
\end{figure}
\begin{figure}[t]
    \includegraphics[width=8.6cm]{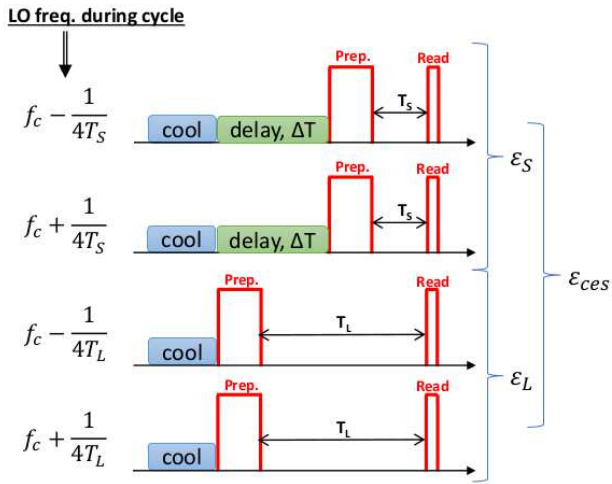}
    \caption{
    \label{Figure2_FJCES_sequence}
    The experimental sequence for CES spectroscopy with a cold-atom CPT clock. Two sub-sequences, each comprised of two Ramsey-interrogations, are applied. Each interrogation begins with a cooling period. For clock interrogations with a short dark-period $T_S$, a delay $\Delta T=T_L-T_S$ is introduced between the end of the cooling phase and the beginning of the Ramsey interrogation. 
    }
\end{figure}

Atomic clocks are exquisite instruments that enable high-performance precision measurements with unrivaled stability and accuracy up to the 10$^{-18}$ range \cite{SchioppoNatPhoton2017}. With their extreme sensitivity, atomic clocks are poised to play an important role in a wide variety of fundamental research activities including relativistic and chronometric geodesy \cite{GrottiNatPhys2018}, the measurement of possible variations of fundamental constants \cite{NemitzNatPhoton2016, LisdatNatComm2016, SfaranovaPRL2018}, the search for dark matter in the universe \cite{DereviankoNatPhys2014}, and the detection of gravitational waves \cite{KolkowitzPRD2016}.\\
In many atomic clocks, the interaction of atoms with the probing field perturbs the atomic energy levels and induces a systematic frequency shift of the clock transition that limits the ultimate clock frequency accuracy and stability. These limitations are of major concern in several types of atomic clocks, including compact microwave atomic clocks based on coherent population trapping (CPT) \cite{HemmerJOSAB1989, YanoPRA2014, BlanshanPRA2015, HafizJAP2017, YunPRappl2017, PollockPRA2018}, as well as optical clocks based on the probing of ultra-narrow quadrupole \cite{HuangPRA2012}, octupole \cite{HosakaPRA2009} and two-photon transitions \cite{BadrPRA2006} or direct frequency-comb spectroscopy \cite{FortierPRL2006}.\\
Ramsey's method of separated oscillating fields \cite{RamseyPhysRev1950} is an elegant technique for measuring atomic and molecular spectra, with which two short interrogation pulses are separated by a dark-period. Although the atoms spend a significant amount of time in the dark, this approach suffers from a non-negligible residual sensitivity to frequency shifts induced during the interrogation or detection pulses.\\
To overcome this problem, Ramsey-based interrogation protocols using composite laser pulse sequences have been proposed and demonstrated to provide a robust immunity of the clock frequency to systematic shifts induced by the interaction pulses \cite{YudinPRA2010,ZanonRPP2018}. Among them, the Auto-Balanced Ramsey (ABR) scheme was developed and applied to an optical Yb$^+$ ion clock \cite{SannerPRL2018} and later demonstrated with CPT-based hot vapor cell \cite{HafizPRAppl2018, HafizAPL2018} and cold-atom microwave clocks \cite{ShukerarXiv2018}. ABR is a powerful approach that utilizes two consecutive Ramsey sequences with different dark periods from which two error signals are extracted and used to control the clock frequency and a concomitant control parameter, thereby compensating for the light-induced frequency shift. A generalized description of the ABR protocol has been reported \cite{YudinPRAppl2018}, which suggests the use of different possible physical variable options as the concomitant parameter.\\
In a recent study, a novel method named combined error signal (CES) spectroscopy has been proposed to form the error signal for the stabilization of Ramsey-based atomic clocks \cite{YudinNJP2018}. Similar to the ABR protocol, the CES method is based on two consecutive Ramsey sub-sequences with different dark periods. However, the CES method uses a single combined error signal, constructed by subtracting the error signals obtained from the two Ramsey sub-sequences with an appropriate normalization factor. The CES method offers light-shift mitigation with a single error signal and a single control parameter, the clock frequency $f_c$. This one-loop method avoids noise and control-related instability associated with two-loop control systems, that are necessarily present with ABR-like protocols.\\
In this letter, we demonstrate the experimental implementation of the CES technique in an atomic clock based on Ramsey spectroscopy. In addition, the error signals are obtained by using frequency jumps (rather than phase-jumps), which prevents any requirement for the modulation or control of the local oscillator's (LO) phase. This results in a very simple implementation of the clock operation, requiring only the control of the LO frequency. \\
The CES method with frequency jumps relies on the fact that the interrogation-related shifts are inversely proportional to the Ramsey dark-period, $T$. The CES sequence is comprised of two sub-sequences, one with a long dark-period $T_L$ and the other with a short dark-period $T_S$. In each sub-sequence, two Ramsey interrogations are performed with the LO frequency set to $f_c+\frac{1}{4T}$ and $f_c-\frac{1}{4T}$. The difference between the transmitted signal in the two cycles is used to compute the error signal associated with this sub-sequence.\\
Fig. \ref{Figure1_CES_principle}.A shows an example of two Ramsey-CPT fringes with $T_L=16$ ms and $T_S=4$ ms, assumed to be light-shifted by $3$ Hz and $12$ Hz, respectively (neglecting all other shifts). Typically, the amplitude of the Ramsey fringe with the longer dark-period is lower due to excess decay. The circles in Fig. \ref{Figure1_CES_principle}.A indicate the sampling points in the case where the frequency jumps are applied from the unperturbed clock frequency, with values of $S_1, S_2$ ($S_3, S_4$) for the long (short) dark-period. Figure \ref{Figure1_CES_principle}.B depicts the error signal $\varepsilon_L=S_1-S_2$ and the normalized error signal $\beta_{cal}\varepsilon_S=\beta_{cal}(S_3-S_4$) extracted from the Ramsey fringes shown in Fig. \ref{Figure1_CES_principle}.A, where $\beta_{cal}$ is a normalization factor. It is evident that both $\varepsilon_L$ and $\beta_{cal}\varepsilon_S$ are non-zero at the unperturbed resonant frequency. On the contrary, the combined error signal $\varepsilon_{CES}$, defined by:
\begin{equation}
\varepsilon_{CES}=\varepsilon_L-\beta_{cal}\varepsilon_S\
\end{equation}
exhibits a zero-crossing at the unperturbed resonant frequency indicated by an arrow in Fig. \ref{Figure1_CES_principle}.B. The normalization factor $\beta_{cal}$ is used to equalize the amplitudes of the long and short fringes, and is given by:
\begin{equation}
\label{Equation_two}
\beta_{cal}=\beta_{decay\,}^{(at)}\frac{N({T_L})}{N({T_S})}.
\end{equation}
Several processes contribute to the fringe amplitude difference, which can be separated into the atomic decay rate, $\beta_{decay\,}^{(at)}$, and the loss of active atoms due to the longer dark-period, $N(T_L)/N(T_S)$, where $N(T)$ is the number of active atoms in a Ramsey cycle with a dark-period $T$. The atomic decay rate $\beta_{decay\,}^{(at)}$ is related to the decoherence or dephasing of the atomic ensemble dark-state, which can be caused by collisions, magnetic field inhomogeneity and other phenomena. In particular, when the atomic decay is dominated by decoherence and loss of active atoms is negligible, eq. \eqref{Equation_two} reduces to\cite{YudinNJP2018} $\beta_{cal}=e^{-\gamma(T_L-T_S)}$, where $\gamma$ is the atomic relaxation rate. The drop in the number of active atoms is a classical effect related to the motion of the atoms and the probing geometry (e.g. size of CPT beam). It is important to stress that an accurate estimation of the normalization factor $\beta_{cal}$ is required to achieve a zero-crossing of $\varepsilon_{CES}$ at the unperturbed resonance frequency, resulting in light-shift cancellation.\\
In the general case, the normalization factor can be obtained by additional Ramsey cycles aiming to measure the peak of the Ramsey fringe \cite{YudinNJP2018}. In the present work, we have applied the CES sequence to a CPT-based cold-atom clock in which the fringe amplitude decay is dominated by atoms escaping the probe region. Since the atomic decay rate is negligible ($\beta_{decay\,}^{(at)}\cong 1$) we were able to achieve normalization simply by introducing a time delay $\Delta T=T_L-T_S$ in the short Ramsey cycle between the turn-off of the cooling light and the beginning of the Ramsey interrogation. Using this approach, the second Ramsey pulse timing is the same for the long and the short Ramsey cycles and the number of active atoms in the two cycles is similar ($N(T_L)/N(T_S+\Delta T)=1$), resulting in an inherent normalization ($\beta_{cal}\cong 1$). In our setup, while using $\beta_{cal}=1$, this delay-based normalization resulted in a fringes amplitude difference of less than $3$ $\%$ (compared with a factor of three difference with no delay).\\
Figure \ref{Figure2_FJCES_sequence} depicts the interrogation sequence for a cold atom CPT clock based on CES. The clock cycle starts with a cooling period (after which the cooling fields are turned off) followed by the Ramsey interrogation. In the Ramsey interrogation, two sub-sequences with a long and a short dark-periods, each composed of two cycles associated with positive and negative frequency jumps are applied to the atoms. The normalization of the fringe amplitude is highlighted by the presence of the time delay $\Delta T$ in the short cycles.\\
The CPT-based cold-atom clock with which we have demonstrated the CES protocol has been described previously \cite{LiuPRAppl2017, LiuAPL2017}. A six-beam magneto-optical trap (MOT) is used to trap and cool about $10^6$ $^{87}$Rb atoms to $10$ $\mu$K. A static quantization magnetic field of $4.4$ $\mu$T is produced in the direction of the CPT beam propagation in order to lift the Zeeman degeneracy. The Ramsey interrogation starts with a $3$ ms preparation CPT pulse, followed by a dark-period of duration $T$, and a final $50$ $\mu$s readout CPT pulse that measures the transmission Ramsey signal. The CPT lin$||$lin configuration is used to detect high-contrast CPT resonances \cite{LiuAPL2017}. The CPT light is sent through the MOT chamber and retro-reflected back by a mirror to reduce Doppler frequency shifts \cite{EsnaultPRA2013}. The CPT light is generated by a distributed Bragg reflector laser tuned on the Rb D1 line at $795$ nm. A electro-optic modulator (EOM) driven at 6.834 GHz by a commercial microwave synthesizer, serving as the LO, generates optical sidebands. The optical carrier and the $-1$-order sideband are used as the two fields for CPT interaction. The CPT field intensity ratio can be changed by adjusting the microwave power driving the EOM and is measured using a Fabry-Perot interferometer. The laser frequency detuning can be changed by tuning the frequency of the RF signal that drives an acousto-optic modulator (AOM). The CES sequence is implemented, and the generated combined error signal $\varepsilon_{CES}$ is used to steer the LO frequency. The LO is referenced to a Hydrogen maser, thereby allowing us to evaluate the clock frequency shift.\\
\begin{figure}[t]
    \includegraphics[width=8.6cm]{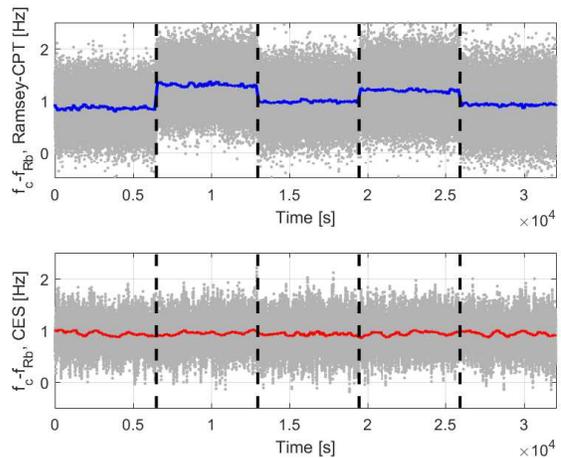}
    \caption{
    \label{Figure3_clock_trace}
    Light shift mitigation with CES spectroscopy. The figure shows traces of the clock frequency (dots are raw data, solid lines are moving averages) for standard Ramsey-CPT case (upper panel) and the CES case (lower panel). During the clock runs, the light shift is changed to different values by jumping the CPT intensity ratio \cite{PollockPRA2018} (dashed vertical lines show the switch times). It is evident that the CES clock frequency remains constant (at a value close to the $2^{nd}$-order Zeemna shift which is $\sim 0.84$ Hz), effectively rejecting the light shifts, whereas the standard Ramsey-CPT clock frequency changes.  
    }
\end{figure}
Figure \ref{Figure3_clock_trace} shows traces of the clock frequency in Ramsey-CPT (upper panel, $T=16$ ms) and CES (lower panel, $T_L=16$ ms, $T_S=4$ ms) protocols. During the clock run, light shifts induced by sudden CPT intensity ratio variations are applied (vertical dashed lines). The traces show the clock frequency $f_c$ subtracted by the generally accepted  $^{87}$Rb ground-state hyperfine splitting frequency, $f_{Rb}$ \cite{RiehleMetrologia2018}. The clock frequency in the Ramsey-CPT case changes abruptly every time the light shift assumes a different value. On the other hand, the clock frequency using the CES method remains nearly constant.\\
\begin{figure}[t]
    \includegraphics[width=8.6cm]{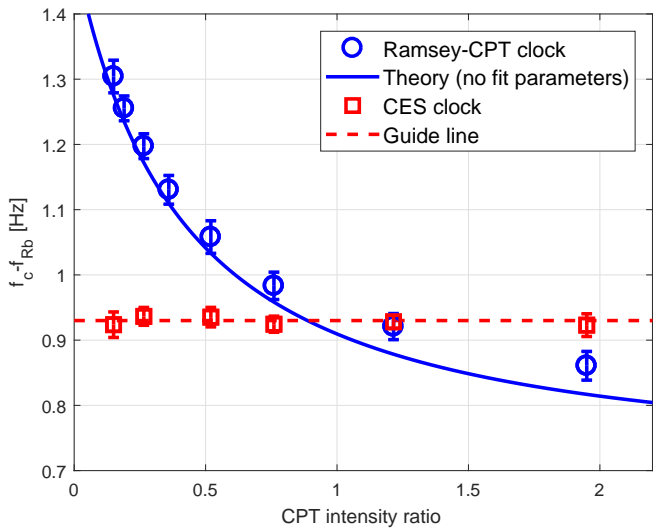}
    \caption{
    \label{Figure4_shifts_averages}
    Measured clock frequency shifts ($f_c-f_{Rb}$) for the Ramsey-CPT (circles) and CES (squares) methods vs. the CPT intensity ratio. In the Ramsey-CPT case, the clock frequency is shifted due to the off-resonant light shift, in good agreement with theoretical predictions \cite{PollockPRA2018} (solid line) with no fit parameters. In the CES case, the clock frequency remains constant within the accuracy of the measurement, thereby eliminating light shifts. We note that the absolute frequency shift of the CES clock is slightly higher than the $2^{nd}$-order Zeeman shift (which is $\sim 0.84$ Hz in this apparatus). We attribute this discrepancy to a residual Doppler shift caused by the alignment inaccuracy of the reflected CPT beam, an effect which is similar for standard Ramsey-CPT and CES protocols. 
    }
\end{figure}
Figure \ref{Figure4_shifts_averages} shows the clock frequency shift versus the CPT intensity ratio in the CES scheme ($T_S=4$ ms, $T_L=16$ ms), in comparison to the Ramsey-CPT scheme ($T=16$ ms). In the Ramsey-CPT case, the experimental results are in excellent agreement with theoretical predictions (with no fit parameters) and are well-explained by off-resonant light shifts \cite{PollockPRA2018}. Through use of the CES interrogation protocol, we observe a reduction of the clock frequency variations by at least an order of magnitude, and the measurements are consistent with complete cancellation of the light shift. These results demonstrate that the CES method significantly reduces the clock frequency sensitivity to the CPT intensity ratio variations.\\
\begin{figure}[t]
    \includegraphics[width=8.6cm]{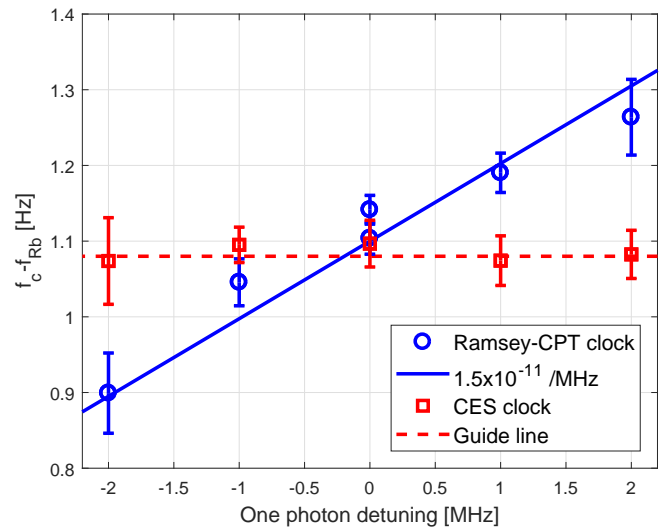}
    \caption{
    \label{Figure5_shifts_averages_OPD}
    Measured clock frequency shifts ($f_c-f_{Rb}$) in the Ramsey-CPT (circles) and CES (squares) methods vs. the one-photon detuning. In the Ramsey-CPT case, the clock frequency changes linearly with OPD due to the resonant light-shift \cite{HemmerJOSAB1989}. The slope is in good agreement with the dependence measured directly from Ramsey fringe spectroscopy ($1.5\pm 0.1\times 10^{-11}$/MHz). In the CES case, the clock frequency remains constant within the accuracy of the measurement and the measured slope is $(0.0\pm0.2)\times 10^{-11}$/MHz. The absolute frequency shift of the CES clock is higher than expected (due to the $2^{nd}$-order Zeeman shift, which is $\sim 0.84$ Hz in this apparatus). We attribute this discrepancy to a residual Doppler shift caused by the alignment accuracy of the reflected CPT beam, which we were able to align to within $\pm0.4$ mrad. 
    }
\end{figure}
In an additional measurement, we have tested the ability of the CES method to reject light shifts caused by one-photon detuning (OPD) of the CPT laser associated with resonant light shifts \cite{HemmerJOSAB1989}. In the Ramsey-CPT case (with $T=16$ ms), the fractional clock frequency dependence on the OPD is $(1.5\pm0.1)\times 10^{-11}$/MHz. In the CES case, the clock frequency remains constant within the accuracy of the measurement, with a fractional clock frequency dependence on the OPD of $(0.0\pm0.2)\times 10^{-11}$/MHz. Similar results were obtained for the $\sigma^+-\sigma^-$ CPT configuration.\\
In conclusion, we have studied the combined error signal (CES) spectroscopy method \cite{YudinNJP2018} and implemented it in a cold-atom CPT clock. The CES method is simple to implement because it uses a single control loop with a single control parameter - the clock frequency. In the current implementation, only the LO frequency needs to be controlled and the fringe amplitude normalization (required in CES spectroscopy) is obtained by introducing a delay between the cooling phase and the Ramsey interrogation equalizing the cycle time of the short and long dark-period cycles. The CES method avoids noise and instabilities associated with more complex methods involving two nested control loops \cite{YudinPRAppl2018}. Our results show a reduction of the light shifts by at least an order-of-magnitude. The CES technique can be applied to a wide range of measurements, including CPT and optical clocks, in order to improve their accuracy and long-term stability. \\
The authors acknowledge helpful discussions with C. Oates and D. Bopp, and thank V. Maurice for help with the implementation of the clock control software. The Russian team was supported by the Russian Science Foundation (No. 16-12-10147). V. I. Yudin was also supported by the Ministry of Education and Science of the Russian Federation (No. 3.1326.2017/4.6), and Russian Foundation for Basic Research (No. 17-02-00570). A. V. Taichenachev was also supported by Russian Foundation for Basic Research (No. 18-02-00822). R. Boudot was supported by the NIST Guest Researcher Fellowship and D\'el\'egation G\'en\'erale de l'Armement (DGA).\\

\bibliographystyle{apsrev4-1}
\bibliography{bib_CACPT}

\end{document}